\documentclass[11pt]{article}

\usepackage{epsfig}
\usepackage{graphicx}
\usepackage{amsmath}
\usepackage{caption2}
\usepackage{bbm}

\begin{document}

\begin{center}
\vskip 0.5 cm
{\large \bf
Dynamical ``breaking'' of time reversal symmetry  and converse quantum ergodicity
}

\vskip 0.5 cm

 BORIS GUTKIN
\vskip 0.3 cm

{\em Mathematisches Institut 
der Universit{\"a}t Erlangen-N{\"u}rnberg,\\ 
Bismarckstr. 1 1/2, D-91054 Erlangen, Germany}\\
{\small E-mail:gutkin@mi.uni-erlangen.de}

\end{center}
\vskip 1.0 cm

\begin{abstract}

\noindent 
It is a common assumption that  quantum systems with  time reversal invariance and classically chaotic dynamics  have   energy spectra distributed according to  GOE-type of statistics.  Here we present a class of systems which fail to follow this rule. We show that for   convex  billiards of constant width  with time reversal symmetry and ''almost`` chaotic dynamics the energy level distribution is of GUE-type. The effect is due to the lack of ergodicity in the ``momentum'' part of the phase space and, as we argue, is generic in two dimensions. Besides, we show that certain  billiards of constant width in multiply connected domains are of interest in relation  to  the quantum ergodicity problem.  These billiards are quantum  ergodic, but not classically ergodic.

\end{abstract}

\vskip 2.0 cm

\section{ Introduction}

The famous conjecture of  Bohigas, Giannoni and Schmit (BGS) \cite{bgs} asserts that the  energy levels  of classically chaotic  systems are distributed as eigenvalues of  random-matrix ensembles. Accordingly, the statistics of the energy levels  is universal and  depend  only on symmetries of the system.  
 That means the energy levels distribution for  (spinless) chaotic systems  with time-reversal invariance  should be  close to that of Gaussian Orthogonal Ensemble (GOE). If the  time-reversal invariance is broken  the distribution of the  energy levels follows statistics of Gaussian Unitary Ensemble  (GUE). The
BGS conjecture   have been supported by broad   numerical and experimental evidence. Indeed, for a large number of systems  without additional symmetries spectral statistics are in agreement with the above predictions. 
 This  might  not be true, however,  if  additional symmetries are present.  Examples of symmetric billiards with anomalous spectral statistics were given in \cite{lss}, \cite{kr}. These billiards are time-reversal invariant, but in addition have a  certain  rotational symmetry. As a result, the   statistics  for  a part of their spectra   turns out to be of GUE-type (rather than  of GOE-type).  On the other hand,  anomalous  spectral statistics may  also appear in  systems with broken time reversal invariance.
For example, the energy levels of magnetic billiards with  reflection symmetry are known to be distributed as in  GOE \cite{br1}. 

 In this paper  we introduce a class of  convex billiards of constant width  with smooth boundaries 
whose dynamics  are time-reversal invariant and ``almost'' chaotic. These billiards have no   additional  symmetries  but, nevertheless, exhibit statistics of GUE-type. On the contrary, if an additional reflectional symmetry is present, the spectral statistics turns to be of GOE-type. We then give an elementary semiclassical explanation for these results  and discuss the implications for spectra of generic convex billiards with smooth boundaries.   

In the second part of the paper we deal with fully chaotic billiards of constant width in the multiply connected domains. As we explain below, the   spectral statistics of these billiards  differ from those of smooth convex billiards of constant width. Nevertheless,  it  has been recently shown  in \cite{robnik} that  for large energy ranges  the  spectrum exhibits GUE like behavior, as well. Our interest to these billiards  here, is primely  related to the converse quantum ergodicity problem.  By Schnirelman theorem \cite{schnir} classical ergodicity implies quantum ergodicity. The converse question  make sense too \cite{zelditch}: Are  quantum ergodic systems  necessarily classical ergodic?   As we demonstrate, certain symmetric billiards of the above type  provide  negative answer to this question. In other words, they are quantum ergodic but not classically. To the best of our knowlage this is the first  example of such  systems.

\section{ Billiards of constant width}

We deal  with a class of convex  billiard tables of constant width with smooth boundaries.
That means the maximal distance between any point on the billiard boundary and 
other boundary points  is a constant.  There is a simple way to construct  such  billiards by  using the following   
  parameterization for the billiard boundaries $z(\alpha)$ in the complex plane: 

\begin{equation}
 z(\alpha)= z(0)-i\sum_{n\in Z} \frac{a_n}{n+1}\left ( 
e^{i\alpha(n+1)}-1\right), \qquad \alpha\in [0,2\pi).
 \end{equation}
 Here $ z(\alpha)$  defines  a curve  of   constant width, whenever  the  parameters $a_n$'s   satisfy conditions: $a_{-n}=a_n^*$, $a_1=0$, $a_{2n}=0$ for $n>0$  \cite{knill}.
Previously,  billiards of constant width  have 
attracted an attention   due to their unusual geometry of caustics.  Our interest  here stems from their peculiar dynamical 
property:
a billiard trajectory which hits the boundary with an angle  in the interval 
$[0,\pi/2]$ (resp.\ $[\pi/2,\pi]$) must   hit the boundary at the next bouncing point  with an  angle 
at  the same interval. In other words, the billiard  ball once launched clockwise (resp.  anti-clockwise) will move in that way forever.

 \begin{figure}[htb]
\begin{center}
\includegraphics[width=7.3cm]{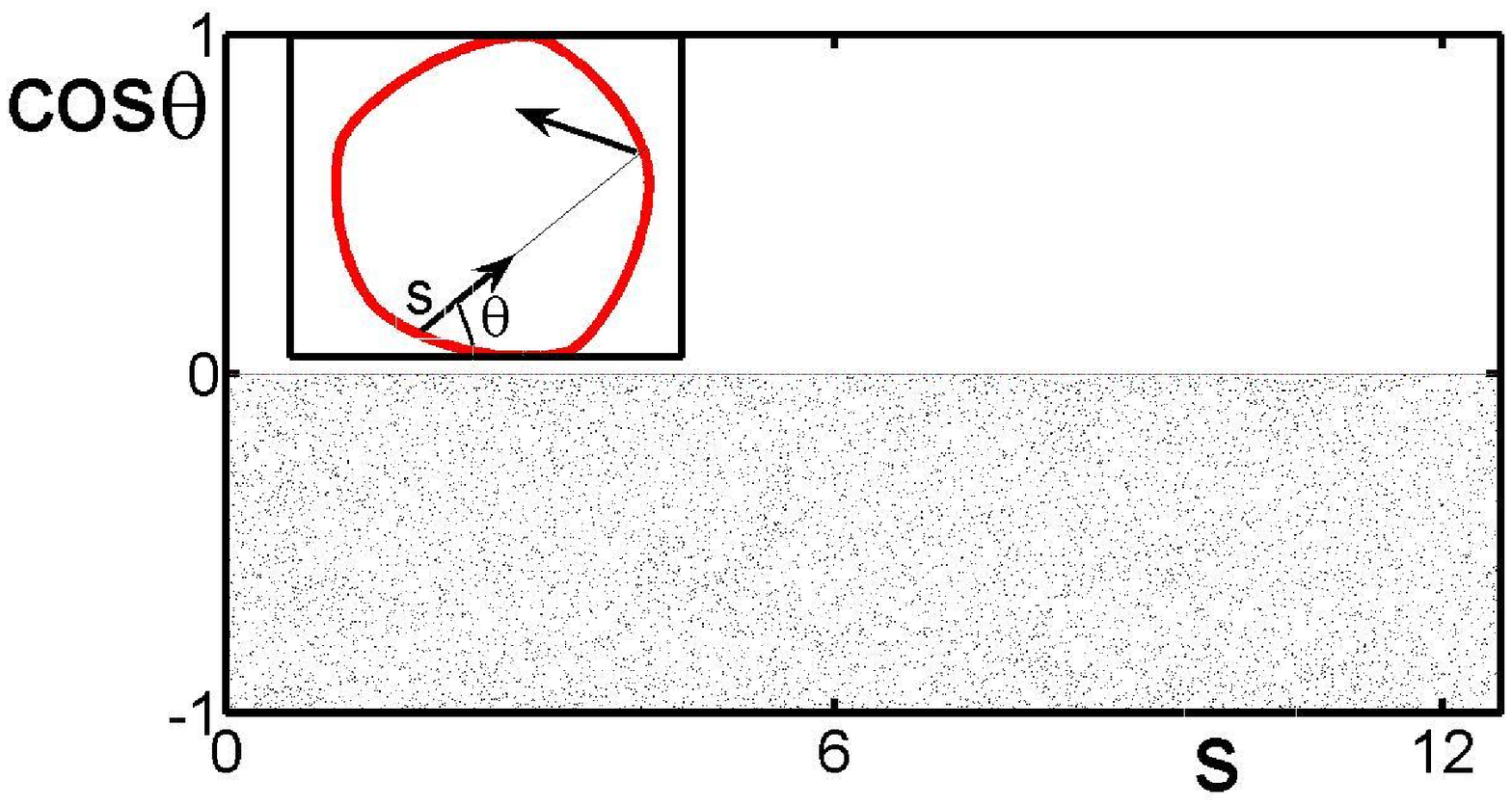}
\hskip 0.1cm
\includegraphics[width=7.3cm]{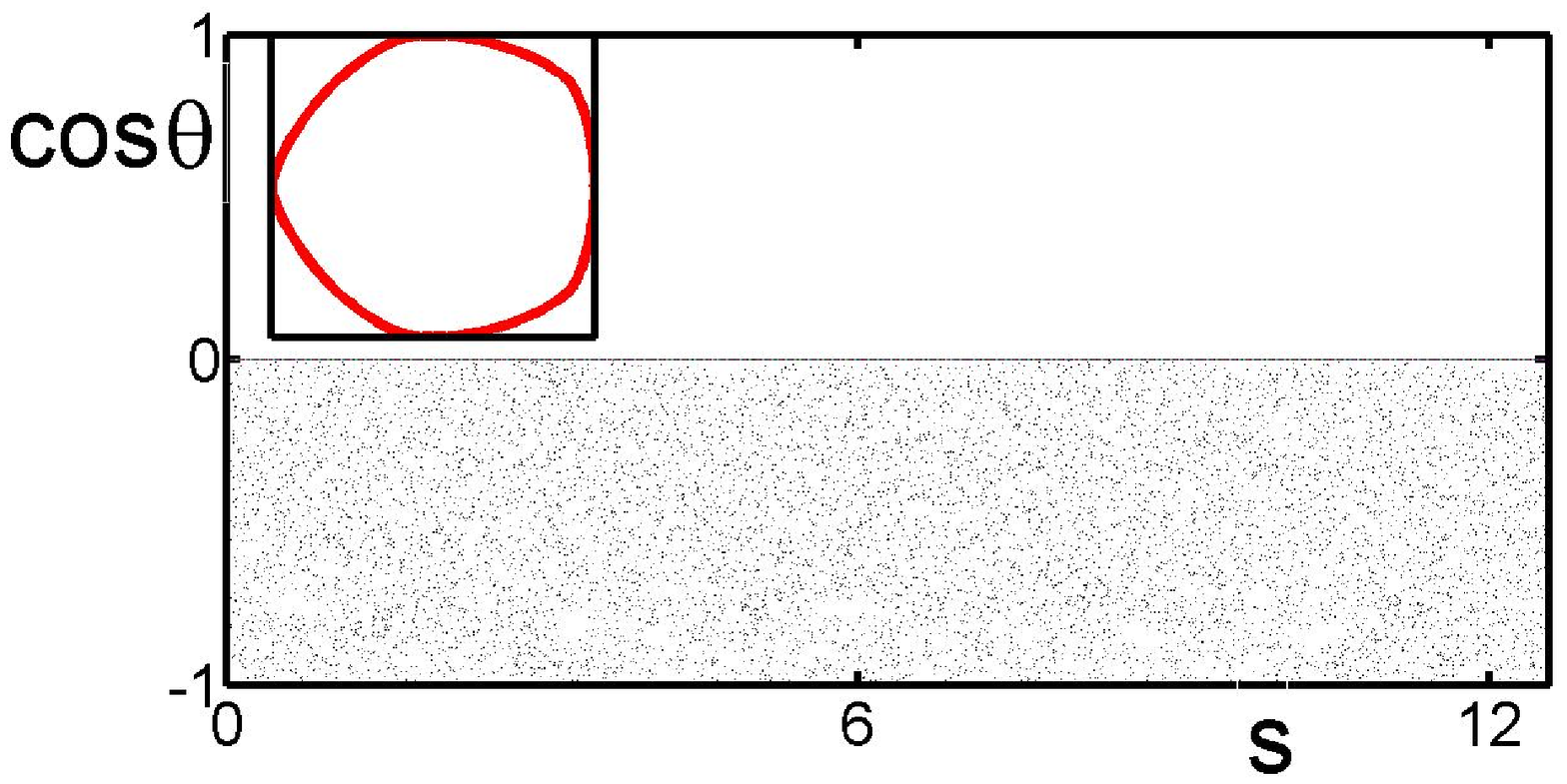}
\end{center}
\caption{\small{ The insets show non-symmetric (left) and symmetric (right)   billiards of constant width  with parameters  $\{a_0=2,a_3=i/4,a_5=1/2+i/2,a_{2k+1}=0,k>2\}$ and $\{a_0=2,a_3=i/4,a_5=-3i/4,a_{2k+1}=0, k>2\}$ respectively. The corresponding  phase space pictures 
are obtained after hundreds of iterations of the billiard map applied  to a number of initial points  located in the lower ($\theta>\pi/2$) half of the phase space. Note, that the iterated points  do not penetrate into the upper half.  This implies  separation of   the clockwise and  anti-clockwise types of motion. (For a comparison with a generic billiard see fig. 4) }}
\end{figure}
The billiard dynamics can be described  in a standard way with the help  of the associated Poincare map. The map acts on   unit vectors attached to the boundary by translating them according to the rules of billiard dynamics. The corresponding two-dimensional phase space can be  parameterized by a couple of  coordinates   prescribing  position and direction of the unite vectors. The canonical choice is  ($s, \cos\theta$), where  $s\in [0, 2\pi a_0)$ is the arclength parameter along the boundary and  $\theta\in [0, \pi]$ is the angle between the unite vector and the tangent line to the boundary. 
In the case of time reversal invariant dynamics the phase space has a reflectional symmetry along the  line $\theta=\pi/2$.  Furthermore, for  billiards of constant width  this symmetry line is invariant under the billiard map and separates  the 
motions in the clockwise and anti-clockwise directions, see fig. 1. 
In general,  a  billiard of constant width defined by eq. 1   has a mixed phase space where regions of regular motion, i.e., Kolmogorov-Arnold-Moser (KAM) tori and 
elliptic islands coexist with regions of chaotic motion. In particular, in the vicinity of the  line $\theta=\pi/2$  there always exists   a region  filled by KAM tori. For our purposes, it 
will make sense to consider billiards  with ``maximally'' chaotic phase space. Two such  billiards    are shown in fig. 1.  Here the 
parameters $a_n$ are adjusted in a way to minimize   the sizes of elliptic islands, 
as well as,  regions of KAM tori along the lines $\theta=\frac{\pi}{2}$ and $\theta=0,\pi$  (whispering gallery  region).

\section{ Spectral statistics}

We consider the energy spectrum of the  quantum billiards in fig. 1 
with the Dirichlet boundary conditions. As has been explained above, the change   from clockwise to  anti-clockwise   types of motion is
classically forbidden process. 
 On the other hand,  in  the quantum billiards such a  switch is possible due to the tunneling effect. The corresponding  
 tunneling  time $\tau$ is finite and depends on 
the width    of  the ``dynamical barrier'' along the separation line $\theta=\frac{\pi}{2}$. This leads to a special structure of the energy spectrum. Most of the  energy levels can be separated into  
quasidegenerate pairs $\{ E^{s}_n, E^{a}_n \}$. Here $E^{s}_n$, $E^{a}_n$ 
correspond to the symmetric $\varphi^{s}_n\approx (\varphi^{l}_n+\varphi^{r}_n)/\sqrt2$ and  the antisymmetric $\varphi^{a}_n\approx (\varphi^{l}_n-\varphi^{r}_n)/\sqrt2$  combinations of the clockwise $\varphi^{r}_n
$ and the
anti-clockwise $\varphi^{l}_n
$ quasimodes. The Wigner transforms of $\varphi^{r}_n
$, $\varphi^{l}_n
$ are  entirely concentrated in the ``lower'' ($\theta<\pi/2$)  and the ``upper'' ($\theta>\pi/2$) half of the phase space, respectively. Furthermore, the splittings of the energy
 levels $\delta E_n=|E^{a}_n-E^{s}_n|$ determine the  time 
$\tau \sim \hbar/\left\langle \delta E_n \right\rangle $ 
needed for a wavepacket to switch the direction of motion. Since the tunneling time  through the ``dynamical barrier'' is   exponentially large in $\hbar$,  this results in  exponentially small splittings $ \delta E_n$ between the energy levels. 
The rest of the spectrum \{$E^0_n$\} contains  eigenfunctions  $\varphi^{0}_n$ whose Wigner transform
 localized    in the  invariant neighbourghood of the line $\theta=\pi/2$. For  the considered range of energies the  only visible part of  \{$E^0_n$\} are  
unpaired zero angular momentum bouncing modes localized  in the phase space exactly on the separation line.

 Since all  paired levels $ (E^{s}_n, E^{a}_n )$ are quasi-degenerate it makes sense to consider  half of the spectra, e.g., $E^{s}_n$.    
(The levels $E^{0}_n$ constitutes a tiny fraction of the  whole spectrum   and have no significant  impact on the spectral statistics   anyway.) Our  primary interest  is the spectral statistics of  the
energy levels $E^{s}_n$. To this end we have numericaly 
calculated by the scaling method of Vergini and Saraceno \cite{vergini} a number ($\sim 15000$) of the energy levels for each of the billiards in fig. 1. The 
results for the nearest-neighbouring distribution $P(s)$ are peresented  in fig. 2. As one 
can clearly see, the distribution for the   billiard without additional  symmetries (left in fig. 1) clearly follows the pattern of GUE. This contradicts a common belief that chaotic systems with time reversal invariance  have spectra  of GOE type   when  additional symmetries are absent. In contrast,  the distribution $P(E)$ for the  billiard with a reflectional symmetry (right in fig. 1) exhibits GOE type of statistics. Below we provide an elementary  explanation for these results based  on the semiclassical link between  spectral statistics and 
periodic orbits of the system. Specifically, let us focus on the spectral
form factor $K(T)$.  It  is defined as the Fourier transform of the autocorrelation 
function 
\begin{equation}R(s)=\bar{d}^{-2}\left\langle d( E+s) d(s) \right\rangle-1, \end{equation}
where $d(E)= \sum \delta(E-E_n)$, $\bar{d}=\left\langle d(E) \right\rangle$ denote the density of states and its mean value. By means of 
the semiclassical trace formula the density of states can be written  as a sum $d(E)= \bar{d}+\sum A_n \exp(iS_n(E)/\hbar)$ over periodic orbits, 
where  phases $S_n(E)$ include both actions and Maslov indices of the periodic orbits. After substitution of $d(E)$  into  (2) and taking the Fourier transform one gets the semiclassical representation of 
$K(T)$ as   double sum over pairs of periodic orbits. The spectral form factor can be naturally separated $K(T)=K_{diag}(T)+K_{off}(T)$ into
 two  terms  provided by  diagonal  ($S_i=S_j$) and off-diagonal  ($S_i \neq S_j$)  correlations of periodic orbits.
\begin{figure}[htb]
\begin{center}
\includegraphics[width=7.3cm]{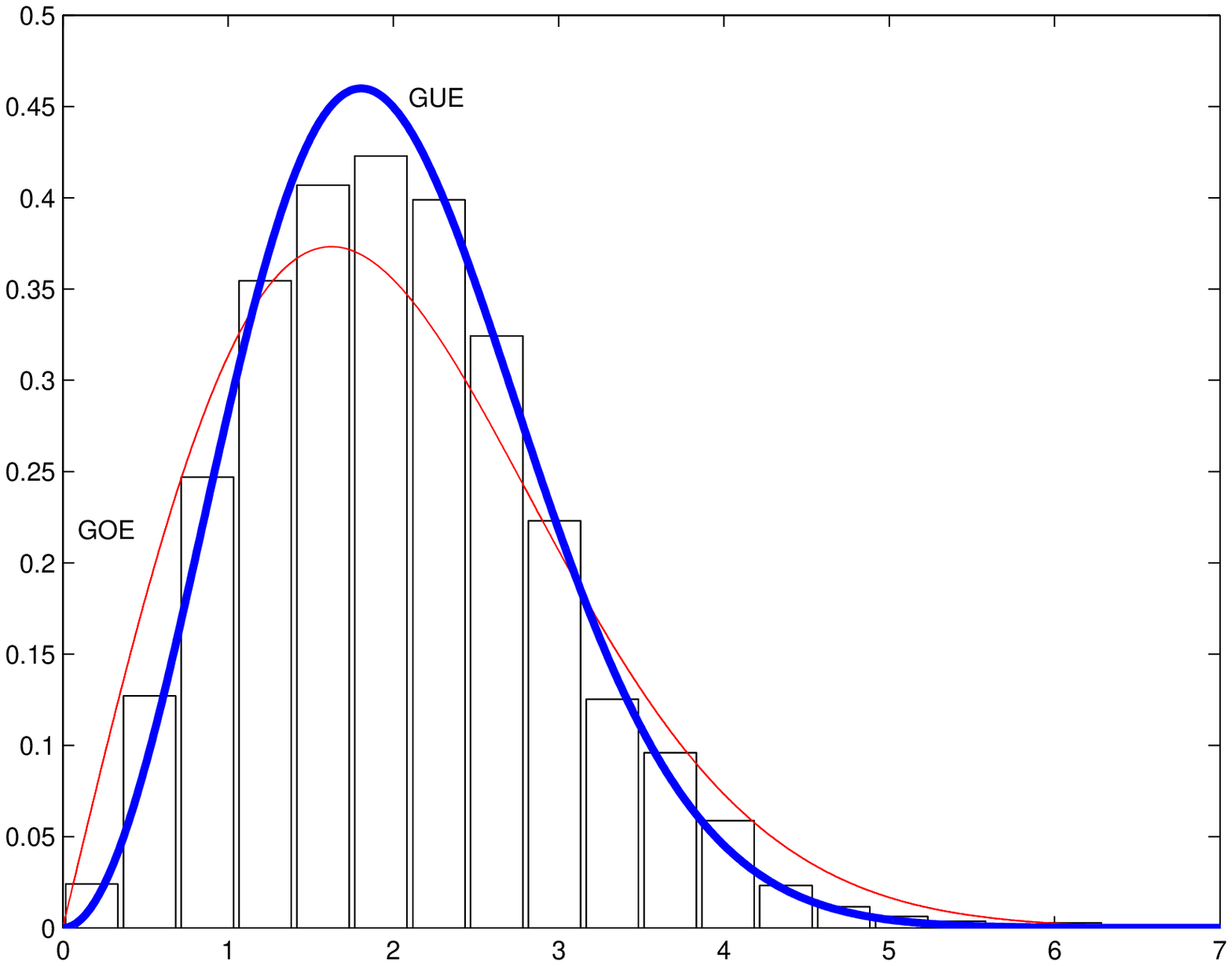}
\hskip 0.1cm
\includegraphics[width=7.3cm]{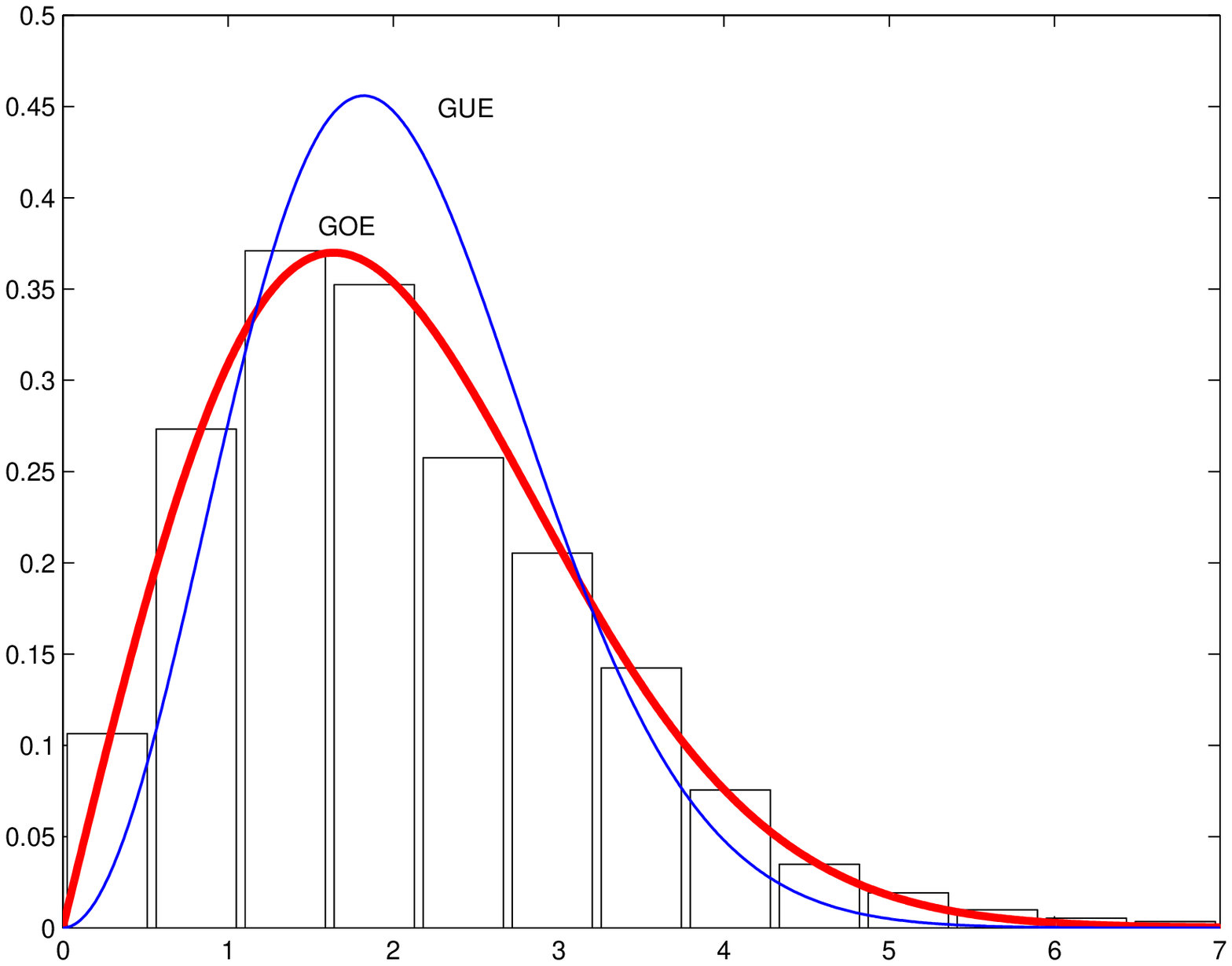}
\end{center}
\caption{Nearest neighbor distribution of energy levels for  nonsymmetric (left) and symmetric (right) billiards in fig. 1. For comparison, GOE and GUE predictions are shown as well.}
\end{figure}
 The leading diagonal term  was derived  by Berry \cite{berry} and in   the Heisenberg time $T_H=2\pi\hbar\bar{d}$ units $t=T/T_H$ found to be  $K_{diag}(t)=\beta  t$, with 
$\beta=2$ for time reversal invariant systems  and $\beta=1$ otherwise.
This should be compared with the spectral form factors 
\[K_{GUE}(t)=t, \qquad  K_{GOE}(t)=2t+t\ln(2t+1),  \mbox{ for }  t<1  \] 
for GUE and GOE.  
In the absence of time invariance $K_{off}$ vanishes and the diagonal term alone reproduce correctly $K_{GUE}$. On the contrary, for  time reversal invariant systems $K_{diag}$ gives  only leading term and  the off-diagonal correlations between periodic orbits must provide  the rest. 
  It has been first shown in \cite{rs} that  GOE result can   indeed be  reproduced   correctly if one takes into consideration the correlations 
between pairs of  self-encountered  
periodic trajectories which  approach themself  from the 
opposite directions under small angles.   As a result, the term of order $n>1$ in the Taylor expansion of $K_{GOE}(t)$  
comes from the correlations of pairs of periodic orbits with $n-1$ 
self-encounters. It is a straightforward observation that  pairs of self-encountered periodic orbits just do not exist in  
billiards of constant width, since trajectories cannot reverse their directions of motion, see fig. 3. This implies that  $K_{off}(t)$ must  be zero and 
$K(t)=K_{GUE}(t)\equiv t$. Hence, the spectral form factor  of the  non-symmetric  billiard in fig. 1 (left) should be  of GUE and not of GOE type. 
On the other hand, for  the  billiard in fig. 1 (right)  the reflectional symmetry substitutes the role of time reversal invariance and  restores correlations between periodic orbits.
 (The simplest way to see this is to consider a half of the billiard. The dynamics there  are not ``uni-directional`` and self-encountered trajectories do exist.) That leads back to GOE type of spectral statistics. This is in complete analogy with the case  of  reflectional symmetric billiards in the presence of magnetic field, where one observes GOE statistics instead of GUE \cite{br1}.

\begin{figure}[htb]
\begin{center}
(A) \includegraphics[height=2.2cm]{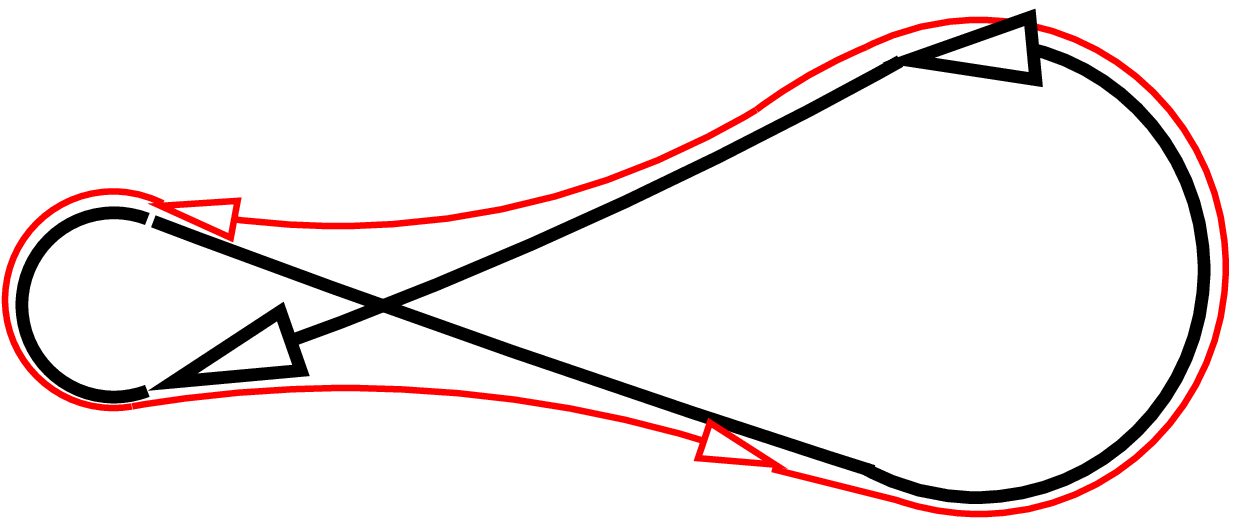}
\hskip 2.0cm
(B) \includegraphics[height=2.2cm]{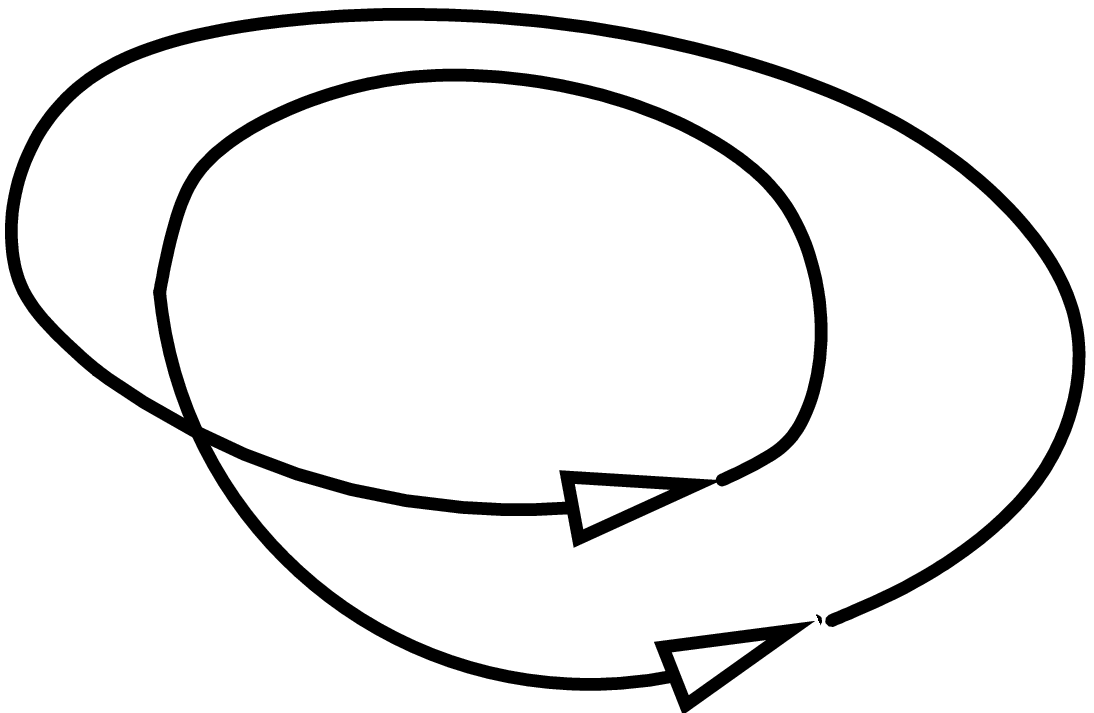}
\end{center}
\caption{{\small Sketch of a pair of  self-encountered periodic  orbits (A), and   of  ``typical''  periodic orbit in a  billiard of constant width (B). Note that  pairs of  self-encountered periodic  orbits do not appear in  billiards of constant width.} }
\end{figure}

\section{Generic convex billiards}

Let us consider now the implications of the above results  for  spectra of  generic billiards whose dynamics  is neither fully integrable nor chaotic. In that case, by the Berry-Robnik theory  \cite{br2} the  energy  spectrum   is 
composed of the independent spectra corresponding to the invariant 
parts of the phase space. Thus for systems with time reversal symmetry, one 
might  expect the  energy level distribution be a mixture of the Poissonian statistics  with   GOE type statistics  related to   regular and 
 chaotic parts of dynamics, respectively. However, as we argue below, a general picture could be somewhat more complex. For a typical billiard with smooth boundaries the chaotic part of the phase space  could be separated into two types of invariant regions, as shown in  fig. 4. In the   regions of the first type the dynamics are   bi-directional. That means   the billiard ball launched from such a region  might  reverse the  direction of flight  in the course of motion.  As a result, the periodic trajectories  admit self-encountering and one can expect  the corresponding statistics be of GOE type. In  the regions of the second type  the dynamics are always uni-directional. Here the full switch of flight  direction is prevented by KAM tori and the resulting statistics should be of GUE type. Thus, in the absence of spatial symmetries the overall  spectral statistics must be a mixture of independent GUE, GOE  and Poissonian statistics   corresponding to the invariant sets with uni-directional, bi-directional chaotic dynamics  and  regular dynamics. If an additional reflectional symmetry exists in the billiard  then only GOE and Poissonian parts are present. This should describe  a typical  structure of spectra for  two dimensional billiards with smooth boundaries. In more  then two dimensions, however, KAM tori do not  separate  regions of phase space. So it could be expected that  typically only bi-directional  type of motion exists and only GOE type of  subspectra  appear.   

   \begin{figure}[htb]
\begin{center}
\includegraphics[width=12.5cm]{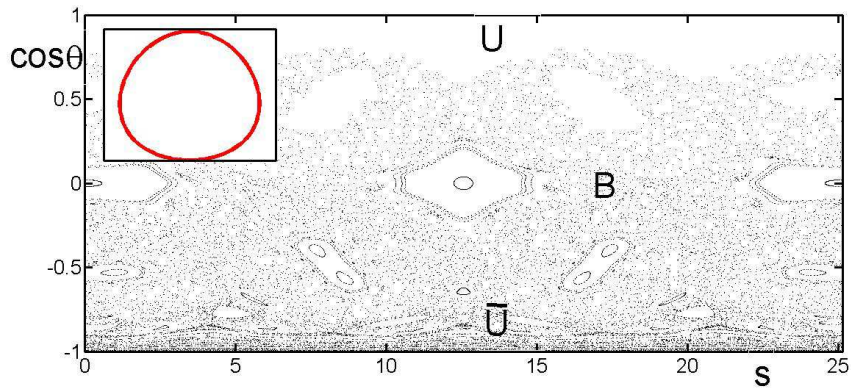}
\end{center}
\caption{A ``generic'' billiard ($a_2\neq 0$) with parameters  $a_0=4, a_2=0.1, a_3=0.5, a_5=0.1,  a_k=0, k>5$ defined by eq. 1. The billiard map is applied to the initial points located in the lower  half of the phase space. Note, that the iterated points do not penetrate into a certain  domain $U$. Hence,  $U$ is dynamically separated  from the  symmetric domain $\bar{U}$ and the corresponding dynamics are ``uni-directional''.  On the contrary, the dynamics in the central part $B$ of the phase space are ``bi-directional''.}
\end{figure}

It worth to  notice that a  mixture of large number of independent spectra would result in the Poissonian statistics, irrespectively of the statistics of individual components.  Thus for generic systems it  would be hard, in practice,  to see  the appearance of GUE type subspectra. The  billiards of constant width represent a very special class of dynamical systems where  bi-directional type of dynamics is completely absent and  the effect can be clearly observed.

\section{Fully chaotic billiards and converse quantum ergodicity} 

The billiards considered so far are only ``approximately''   chaotic. But it is also 
possible to  construct  fully chaotic  billiards of constant width  in multiply connected domains. An example  of ``hippodrome'' like billiard  is shown on fig. 5a. By Wojtkowski's cone field method \cite{wojtkowskij} it can be easily shown that  the dynamics in this  billiard   are 
fully hyperbolic  with a positive Lyapunov exponent almost everywhere. Note, however that the billiard  is  not ergodic, since there are  exist (at least) two  ergodic components corresponding to the clockwise and anti-clockwise types of motion separated by the invariant line. Despite of  similar  phase space structure, the quantum spectral properties of the billiards in fig. 1 and fig. 5 are  essentially different. For fully chaotic billiards  the width of separation region between two types of motion shrinks to zero and the tunneling time  is determined by   diffraction effects at the points of  billiard  boundary with curvature jumps. This results in   much shorter  (algebraic rather than exponential in $\hbar$) tunneling times. Thus instead of quasi-degeneracies, one can expect  splitting between  the energy levels be comparable with the mean level spacing.  The  spectral statistics  of  non-symmetric billiards  of such   type (``Monza billiards'') have been recently investigated in \cite{robnik}. It has been shown there, that the long range correlations among levels tend to exhibit  GUE type behavior. This result is in agreement with our observations for convex billiards of constant width  with smooth boundaries.  
 In what follows, we will show that  the ``hippodrome'' like  billiards are also of interest  in connection  with the quantum ergodicity problem.

It is well known that classical ergodicity implies quantum. But is the converse  also true \cite{zelditch,zelditch2}:
 Are quantum ergodic systems necessarily classically ergodic? As we show below the ``hippodrome'' billiard  provide a counterexample.  This billiard is classically not ergodic. Let us show, however, that it is quantum ergodic. First, observe  
that by desymmetrization procedure the full spectrum of the billiard in fig. 5a can be decomposed into the spectra of four  ``quarter'' billiards shown in fig. 5b. The  ``quarter'' billiards have four different combinations $\cal{S}=\{\mbox{DD},\mbox{DN},\mbox{ND},\mbox{NN}\}$ of Neumann and Dirichlet boundary conditions at  two flat pieces of the boundary. We will use index $\nu\in\cal{S}$ to denote  these  billiards and call
$\varphi^{\nu}_n$  the  corresponding eigenfunctions. By definition, any  eigenfunction  $\varphi_{n}$ of the full billiard is just  collection  of four copies of $\varphi^{\nu}_{i}$ for some $\nu$ and $n=\eta_{\nu}(i)$ with $\eta_{\nu}(\cdot):\mathbbm{N}\to\mathbbm{N}$ being the function which maps indices of $\nu$'s quarter billiard eigenfunctions to the indices of the corresponding eigenfunctions of the  full billiard. Note, that  
 each of the desymetrized billiards is classically ergodic  (assuming that there exist only two ergodic components in full billiard). Hence, by Schnirelman's theorem  all four  quantum billiards have eigenfunctions ``equidistributed'' in the phase space. More precisely, this means  for each combination of boundary conditions $\nu$ there exists a  subset of integers $\{j^{\nu}_n\}_{n\in \mathbbm{N}}$ of counting  density one such that  the classical average of  an observable $A(x)$ coincides with the corresponding quantum limit:

\begin{equation}\lim_{n\to\infty}\langle \varphi^{\nu }_{j^{\nu}_n}, \mbox{ Op}\left( A \right) \varphi^{\nu }_{j^{\nu}_n} \rangle =\int A(x) \, d\mu,\end{equation}
where $\mu$ stands for normalized Liouville measure in the   billiard phase space and $\mbox{ Op}( A)$ means Weyl quantization of $A$.
Now, let $V$  be  the phase space  of the full billiard and let $V=\cup_{k=1}^{4}V_k$ be its  decomposition into four symmetric  pieces.    For an observable $O$ in $V$ let  ${O}_k$ denote its restriction to the quarter $V_k$, i.e.,  ${O}_k(x)=  {O}(x)$ if $ x \in V_k$.
 Then  for the corresponding quantum averages one has decomposition 

\begin{equation}\langle \varphi_l, \mbox{Op}\left(O\right) \,\varphi_l \rangle =
\sum_{k=1}^{4}\langle \varphi^{\nu}_{n}, \mbox{Op} \left({O}_k \right)\varphi^{\nu }_{n} \rangle, \qquad l=\eta_{\nu}(n),\end{equation}
where  ${O}_k$ in the righthand side of eq. 4 should be understood  as  observables in the phase spaces  of the  quarter billiards. Now   define the  sequence of integers $\{j_n\}=\cup_{\nu} \{\eta_{\nu}( j^{\nu}_n)\} $. This sequence is of density one and  it follows immediately
from eqs. 3,4: 
\begin{equation}\lim_{n\to\infty}\langle \varphi_{j_n}, \mbox{Op}\left(O\right)  \varphi_{j_n} \rangle =\sum_{k=1}^{4} \int_{V_k}{O}_k \,d\mu=  \int_{V}O\,d\mu,\end{equation}
  implying quantum ergodicity for   classically non-ergodic billiard in fig. 5a.

\begin{figure}[htb]
\begin{center}
(A)
\includegraphics[height=3.0cm]{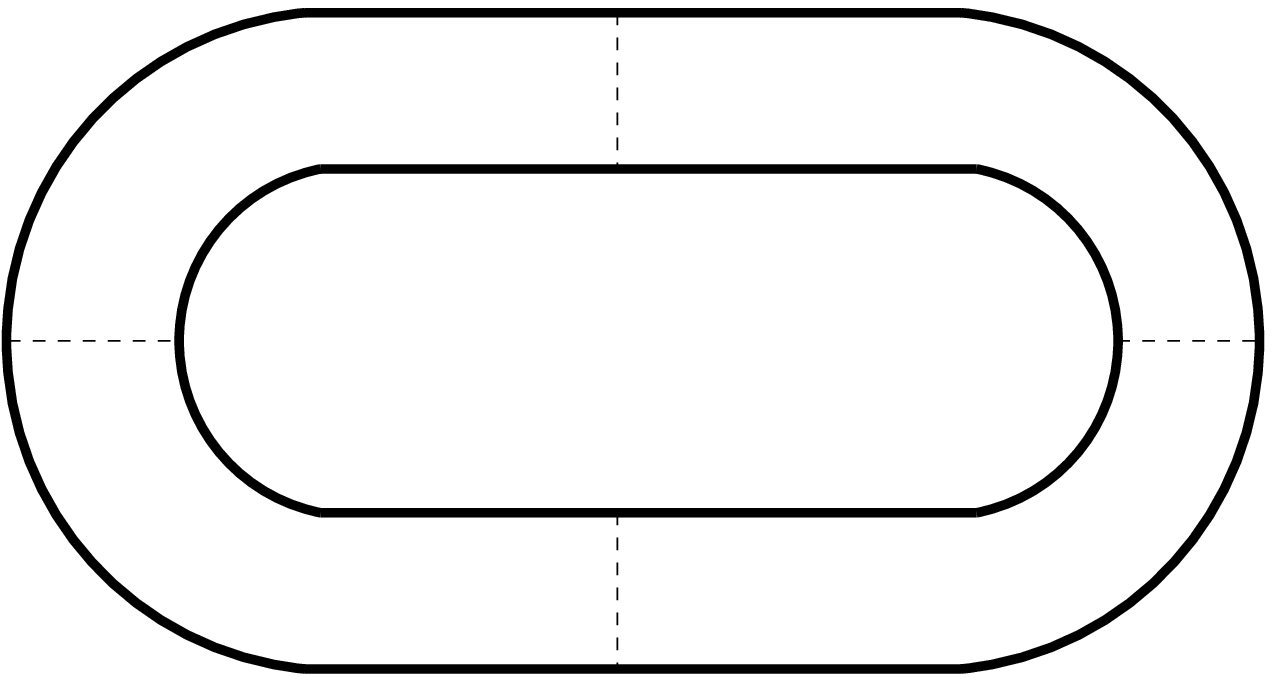}
\hskip 0.5cm
(B)
\includegraphics[height=3.0cm]{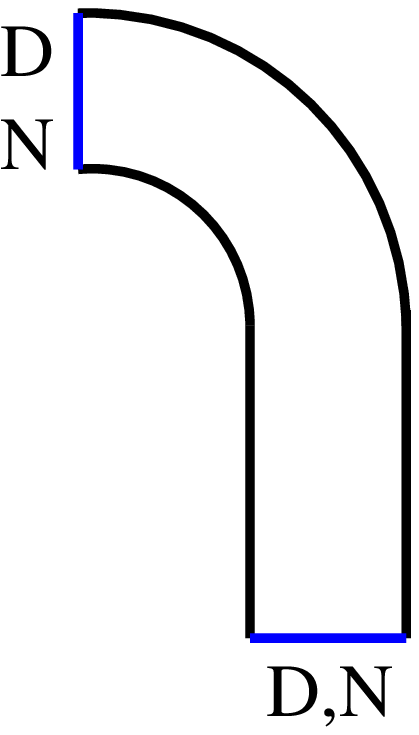}
\hskip 2.0cm
(C)
\includegraphics[height=3.0cm]{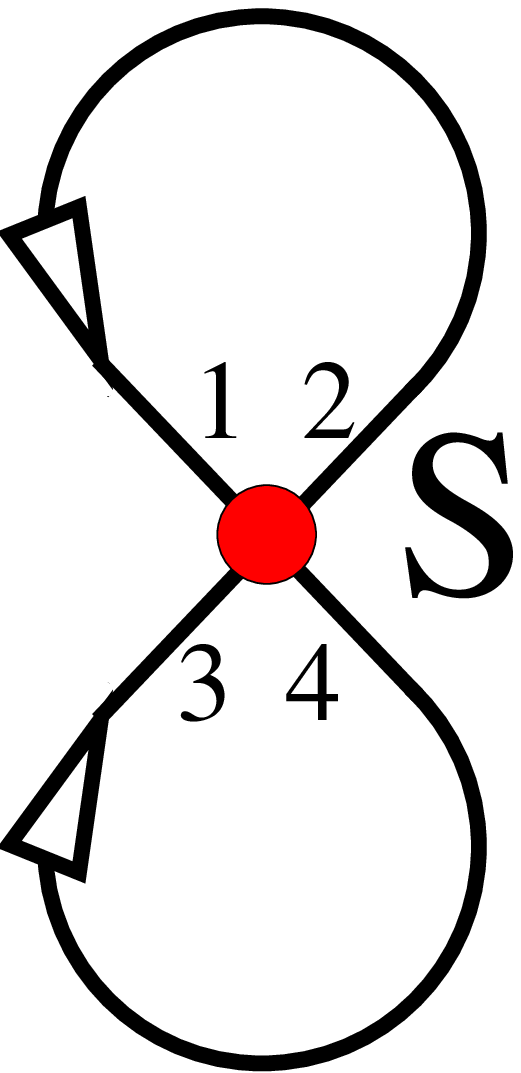}
\end{center}
\caption{\small{On the left: Fully chaotic ``hippodrome'' billiard (A) and its desymmetrized quarter (B) with different combinations of  boundary conditions. Both the external and internal   billiard  walls  are composed of the boundaries of   stadia in a way that  the billiard interior has a constant width. On the right (C): ``Uni-directional'' quantum graph. The scattering matrix at the vertex satisfies conditions: $S_{1,3}=S_{3,1}=S_{2,4}=S_{4,2}=0$, $S_{i,i}=0$,  $i=1,\dots 4$} }
\end{figure}

Let us briefly comment on the reasons for failure of classical ergodicity in the ''hippodrome'' like  billiards. There is a general result stating  two necessary and sufficient conditions for a billiard to be classically ergodic. Let $\{\lambda_n=\sqrt{E_n}/\hbar,   \varphi_{n}\}$ be the billiard spectral data and $N(\lambda)$ be the corresponding counting function, then:

\

\noindent {\bf{Theorem 1}} [See \cite{zelditch2}  and references there.] The billiard flow is ergodic if and only if for every observable $A$:\\

 {\it Condition 1:   } $\lim_{\lambda\to\infty}\frac{1}{N(\lambda)}\sum_{ \lambda_n\leq\lambda}\left| \langle \varphi_{n}, \mbox{Op}\left( A \right) \varphi_{n}  \rangle -\int A(x) \, d\mu\right|^2=0. $ \\

{\it Condition 2:    } $\forall \epsilon, \exists \delta, \,\, \lim_{\lambda\to\infty}\sup\frac{1}{N(\lambda)}\sum_{ \substack{n\neq k, \, \lambda_n,\lambda_k\leq\lambda \\  |\lambda_n-\lambda_k|<\delta}}
\left|\langle \varphi_{n}, \mbox{Op}\left( A \right) \varphi_{k}  \rangle\right|^2<\epsilon. $ \\

\noindent The first condition is essentially equivalent to the quantum ergodicity and as has been  already shown is  satisfied for the hippodrome billiard. Thus the second condition must fail. It is easy to understand the reason for that. Although there are no exponentially small degeneracies in the spectra of  the billiard (as for convex billiards of constant width) its eigenfunctions could be still approximately separated into the pairs of symmetric and antisymmetric combinations:

\begin{equation} \varphi^s_{n}=\frac{1}{\sqrt{2}}(\varphi^r_{n}+ \varphi^l_{n})+r_1, \qquad \varphi^a_{n}=
\frac{1}{\sqrt{2}}(\varphi^r_{n}- \varphi^l_{n})+r_2,\end{equation}
of clockwise $\varphi^r_{n}$ and anti-clockwise $\varphi^l_{n}$ moving quasimodes. Consider now the observable $\chi$ which is  one for points of the phase space  moving in the clockwise direction and zero otherwise. If the reminder terms  $r_1, r_2$ are sufficiently small (as has been numerically observed in \cite{robnik}), the    off-diagonal elements 

\[ \left|\langle \varphi^s_{n}, \mbox{Op}\left( \chi \right) \varphi^a_{n}  \rangle\right|^2=\left|\frac{1}{2}+\frac{1}{\sqrt{2}}\left(\langle \varphi^r_{n},r_2\rangle + \langle r_1,\varphi^r_{n}\rangle\right)+\langle r_1,\mbox{Op}\left( \chi \right) r_2\rangle\right|^2 +O(\lambda^{-1}) > C,\]
are bounded from below, and as a result   Condition 2 is not satisfied. 

\section{Conclusions} 

There are two general  conclusions which can be drawn from the present  study of quantum billiards of constant width. The first one is that in the absence of additional symmetries,   time  reversal invariance of a chaotic system does not automatically guarantees GOE type statistics for the energy spectrum.  
In addition,  the following dynamical condition must be satisfied. Call an invariant ergodic component  $D_i$ of  the  phase space ''time reversally connected'' if for (almost)
 every point $(q,p)\in D_i$  with the  coordinate $q$  and momentum $p$, the ``reverse'' point $(q,-p)$   belongs to  $ D_i$ too. Then (in the absence of additional symmetries) the energy spectrum associated with a  chaotic invariant domain $D$ is of GOE type only if $D$ is ''time reversally connected''. Otherwise, the associated spectrum should be of GUE type.
Loosely speaking, in the systems, such as   billiards of constant width
time reversal symmetry is broken dynamically, rather than with  an external (e.g., magnetic) field.  
Beyond the   spectral statistics,  other quantum properties  in these systems should be affected too.  For instance,  the semiclassical treatment  of the Landauer  conductance in \cite{conduct} and short noise in \cite{noise} through ballistic devices rely on  calculations of the  correlations between selfencounted  periodic trajectories.
So one can use exactly the same arguments as above to derive  GUE  type results for  quantum dots with   non-symmetric shapes of constant width  (resp. GOE for symmetric shapes).

The second conclusion is that classical ergodicity does not follow in general from quantum ergodicity alone. There exist chaotic systems like hippodrome billiards where the diagonal elements satisfy Condition 1 of the Theorem 1 (i.e., quantum ergodicity holds) but off-diagonal terms fail to satisfy Condition 2. This shows that in general some sort of additional condition on off-diagonal terms is, indeed  necessary to  grantee classical ergodicity.     

Finally, it is worth mentioning that besides  billiards, there exist other  systems with   uni-directional  type of dynamics. For instance, the above billiard construction  can be straightforwardly generalized to get  a family of smooth ``uni-directional'' potentials.     Namely,   fix the  coefficients $a_n, n>0 $ in eq. 1, set   $z(0)=-ia_0$ and let  $a_0$ vary over an interval $\Delta=[\delta_1, \delta_2]$, $\delta_{1,2}>0$. For an appropriate choice of $\delta_1, \delta_2$ (that means $\delta_{1}$ cannot be too small) this defines   a family of closed convex curves $\Gamma(a_0), a_0\in \Delta$ of constant width in the  domain  bounded by $\Gamma(\delta_1)$ and  $\Gamma(\delta_2)$.
 Now, let  $V(z)$ be a smooth potential  which is equal  $\infty$  outside  $\Gamma(\delta_{2})$, $0$ (or $\infty$) inside $\Gamma(\delta_{1})$  and  whose  equipotential lines coincide with $\Gamma(a_0), a_0\in \Delta$ in between.  Any such potential gives rise to the uni-directional Hamiltonian flow inside  the domain  bounded by $\Gamma(\delta_{2})$. 
Another class of uni-directional systems is provided by  quantum graphs of a certain type. A  simple  example  is shown in fig. 5c. Here the full separation of two  types of motion is achieved by putting appropriate scattering matrices at the vertices of the  graph.

\

{ Acknowledgments:} I am grateful to J. Avron and S. Fishman for their support during my stay at  Physics Department of Technion. I also would like to thank S. Nonnenmacher for helpful discussions on converse quantum ergodicity problem.
This work was supported by Minerva Foundation.

\end{document}